\documentclass[preprint]{aastex}

\usepackage{natbib}
\usepackage{amsmath,amssymb,graphicx,soul,color}

\shorttitle{fraction of GRBs with high-energy photons}

\slugcomment{Draft version on January 7, 2013}

\begin{document}

\title{Limits to the fraction of high-energy photon emitting gamma-ray bursts}

\author{ Carl~W.~Akerlof\altaffilmark{1}
\email{akerlof@umich.edu}
and  WeiKang~Zheng\altaffilmark{1}
}

\altaffiltext{1} {Department of Physics, University of Michigan, 450 Church Street, Ann Arbor, MI, 48109-1040, USA}

\shortauthors{Akerlof \& Zheng 2012}

\begin{abstract}
After almost 4 years of operation, the two instruments onboard the \it{Fermi Gamma-ray Space Telescope} \rm have
shown that the number of gamma-ray bursts with high energy photon emission above 100 MeV cannot exceed roughly 9\% of the total number of all such events, at least at the present detection limits. In a recent paper (Zheng et al. 2012c), we found that GRBs with photons detected in the Large Area Telescope (LAT) have a surprisingly broad distribution with respect to the observed event photon number. Extrapolation of our empirical fit to numbers of photons below our previous detection limit suggests that the overall rate of such low flux events could be estimated by standard image co-adding techniques. In this case, we have taken advantage of the excellent angular resolution of the $Swift$ mission to provide accurate reference points for 79 GRB events which have eluded any previous correlations with high energy photons. We find a small but significant signal in the co-added field. Guided by the extrapolated power law fit previously obtained  for the number distribution of GRBs with higher fluxes, the data suggests that only a small fraction of GRBs are sources of high energy photons.
\end{abstract}


\section{Introduction}
A most unexpected feature of gamma-ray bursts (GRBs) became apparent in 1994 with the \it{Compton Gamma-Ray Observatory} \rm (CGRO) detection of an 18 GeV photon associated with GRB 940217 (Hurley et al. 1994). About a half-dozen high energy burst events were detected throughout the CGRO mission but the actual fraction of the total rate was poorly determined (Catelli et al. 1998, Dingus 2003). Prior to the launch of the \it{Fermi Gamma-ray Space Telescope} \rm in 2008, it was possible to anticipate that the LAT instrument would detect more than 200 GRB events per year at energies above 100 MeV (Dingus 2003).  Since the GRB spectral energy distribution at lower energies has been well characterized by a modified power law with peak fluxes at energies of the order of 200 KeV (Kaneko et al. 2006, Goldstein et al. 2012), the existence of photons at energies 10$^4$ times higher is a significant constraint on credible models of the GRB phenomenon (e.g. Band et al. 2009). Thus, there had been some anticipation that the Fermi mission might considerably enrich our knowledge of this aspect of GRB behavior. Such hopes were dampened by the Band paper which was drafted 7 months after the {\it Fermi} launch. With three LAT-detected events and an extrapolation of BATSE spectral data, Band et al. concluded that the high energy detection rate of the LAT would be about one per month. Almost four years of operation has demonstrated that the detection rate of energetic GRBs is closer to 9 per year. One interesting discovery that has emerged is that short as well as long bursts contribute to the population of $>100$ MeV emitting GRBs. To date, six short events have been so identified, roughly comparable to their fraction within the total GRB population (GRB 080905A, 081024B, 090228A, 090510, 110529A and 111117A). This suggests that the high energy photon generation process is a characteristic of relativistic jets, independent of the details of their specific progenitors. Given the paucity of information about energetic bursts, we have embarked on a program to dig out as much information as possible from the available data. This has led to a string of papers describing the discovery of seven high energy GRBs with signatures too faint to be detected by more conventional statistical techniques (Akerlof et al. 2010, Akerlof et al. 2011, Zheng et al. 2012a, 2012b, 2012c, 2012d). The following work delves further into this realm to understand better the extent of the association of high energy events with the parent class of all GRBs.
   
\section{Data selection and analysis}
In our first search of LAT data for faint GRBs (Akerlof et al. 2011), we recognized that the precise localizations of bursts ($\sigma_{PSF} < 5'$) detected by Swift, INTEGRAL and AGILE were powerful constraints for identifying associated high energy photons in the Fermi LAT. More recently, we estimated the event intensity distribution and discovered that within the bounds set by limited statistics, the distribution was well fit by the number of LAT photons per unit area raised to a constant negative fixed power (Zheng et al. 2012c). To probe this empirical result at intensities below those required for reliable single event identification, we realized that the standard field co-adding techniques of optical astronomy would work nicely with fields triggered by Swift and similar instruments. The GRB events were selected in the interval from June 11, 2008, ($Fermi$ launch date) through February 29, 2012 from the Swift catalog\footnote{http://swift.gsfc.nasa.gov/docs/swift/archive/grb$\_$table.html/} for which the Fermi LAT boresight angle was computed from the Fermi spacecraft attitude data\footnote{http://fermi.gsfc.nasa.gov/cgi-bin/ssc/LAT/LATDataQuery.cgi}. 
Since the publicly available LAT data stream was restricted to Pass 7 after 2011 August 6, we used Pass 6 prior to the switchover date and Pass 7 thereafter. Events were included only if they passed the maximum LAT acceptance angle of 74$^\circ$ and no previous GRB correlations had been reported. Although the LAT acceptance drops significantly for boresight angles greater than 65$^\circ$, it is still large enough for burst detections such as reported by the LAT team for GRB 100414A (Takahashi et al. 2010) and GRB 120624B (Vianello \& Kocevski 2012). Swift events were also ignored if the Fermi spacecraft was inside the South Atlantic Anomaly (SAA) or was otherwise inactive. A total of 79 events passed these cuts and are listed in Table 1. For reference later in this paper, this sample is called "All$\_Swift$". Since we were interested in investigating possible correlations with the intensity of bursts at lower energies, three further selections were made, based on the Fermi GBM fluence. The fluence data was obtained from either the GBM \it{Fermitrig} \rm  catalog\footnote{http://heasarc.gsfc.nasa.gov/W3Browse/fermi/fermigtrig.html} or GCN messages (Barthelmy 2000). The former took precedence whenever available. There were 46 events with such GBM data and this set is named the "$Swift\_$GBM" sample. Third and fourth samples, "$Swift\_$bright$\_$GBM" and "$Swift\_$dim$\_$GBM" consists respectively of the 14 events with fluences greater than 10 $\mu erg/cm^2$ within "$Swift\_$GBM" and the complementary set of 32 with fluences below this bound. Since "$Swift\_$GBM" is a subset of "All$\_Swift$" and, in turn, "$Swift\_$bright$\_$GBM" and "$Swift\_$dim$\_$GBM" are disjoint subsets of
"$Swift\_$GBM", these four co-add sets are not statistically independent. Nevertheless, there are correlations to the low energy GRB fluxes that makes comparisons useful. It should be noted that very few Pass 7 GRB events are included in any of these data sets: All$\_Swift$ - 5 out of 79, $Swift\_$GBM - 2 out of 46, $Swift\_$bright$\_$GBM - 1 out of 14 and $Swift\_$dim$\_$GBM - 1 out of 32. Thus, any systematic biases that might arise from somewhat different primary data processing are statistically insignificant.

\begin{deluxetable}{lccrrr}
 \tabcolsep 1.4mm
 \tablewidth{0pt}
 \tablecaption{List of 79 {\it Swift}-triggered GRBs}
  \tablehead{
  \colhead{GRB} & \colhead{UT} & \colhead{MET} &
\colhead{RA}           & \colhead{Dec}
& \colhead{$\theta_{bore}$}
\\
  \colhead{}    & \colhead{}   & \colhead{}    & \colhead{($^{\circ}$)}
& \colhead{($^{\circ}$)} & \colhead{($^{\circ}$)}
  }
\startdata
080810  & 13:10:12 &  240066613 & 356.783 &    0.310 &  60.5 \\
080906  & 13:33:16 &  242400797 & 228.055 &  -80.540 &  36.6 \\
080916B & 14:44:47 &  243269088 & 163.632 &   69.061 &  28.6 \\
080928  & 15:01:32 &  244306893 &  95.061 &  -55.176 &  45.3 \\
081003A & 13:46:12 &  244734373 & 262.376 &   16.566 &  56.4 \\
081008  & 19:58:09 &  245188690 & 279.968 &  -57.433 &  63.9 \\
081011  & 00:28:50 &  245377731 & 220.363 &   33.548 &  66.2 \\
081012  & 13:10:23 &  245509824 &  30.184 &  -17.627 &  62.1 \\ 081016B & 19:47:14 &  245879235 &  14.582 &  -43.536 &  62.7 \\
081029  & 01:43:56 &  246937437 & 346.776 &  -68.179 &  57.8 \\
081101  & 11:46:31 &  247232792 &  95.836 &   -0.112 &  29.2 \\
081102  & 17:44:39 &  247340680 & 331.178 &   52.991 &  51.1 \\
081104  & 09:34:42 &  247484083 & 100.500 &  -54.722 &  32.8 \\ 081109A & 07:02:06 &  247906927 & 330.798 &  -54.711 &  72.9 \\
081118  & 14:56:36 &  248712997 &  82.572 &  -43.305 &  29.9 \\
081126  & 21:34:10 &  249428051 & 323.526 &   48.714 &  16.4 \\
081127  & 07:05:08 &  249462309 & 332.075 &    6.858 &  52.1 \\
081204  & 16:44:55 &  250101896 & 349.773 &  -60.221 &  64.9 \\
081222  & 04:53:59 &  251614440 &  22.748 &  -34.095 &  49.3 \\
090113  & 18:40:39 &  253564841 &  32.067 &   33.436 &  30.4 \\
090117  & 15:21:54 &  253898516 & 164.003 &  -58.249 &  52.4 \\
090129  & 21:07:15 &  254956037 & 269.105 &  -32.793 &  24.4 \\
090407  & 10:28:25 &  260792907 &  68.979 &  -12.684 &  33.9 \\
090422  & 03:35:16 &  262064118 & 294.746 &   40.398 &  29.5 \\
090424  & 14:12:09 &  262275131 & 189.531 &   16.829 &  71.0 \\
090509  & 05:10:03 &  263538605 & 241.422 &  -28.385 &  73.1 \\ 090516A & 08:27:50 &  264155272 & 138.246 &  -11.848 &  11.7 \\
090518  & 01:54:44 &  264304486 & 119.957 &    0.778 &  27.7 \\
090519  & 21:08:56 &  264460138 & 142.317 &    0.190 &  46.7 \\
090529  & 14:12:35 &  265299157 & 212.446 &   24.450 &  23.7 \\
090531B & 18:35:56 &  265487758 & 252.070 &  -36.015 &  22.6 \\
090621A & 04:22:43 &  267250965 &  10.987 &   61.938 &   5.5 \\
090702  & 10:40:37 &  268224039 & 175.888 &   11.501 &  59.9 \\
090708  & 03:38:15 &  268717097 & 154.632 &   26.616 &  54.1 \\
090709B & 15:07:42 &  268844864 &  93.522 &   64.081 &  46.9 \\
090712  & 03:51:05 &  269063467 &  70.097 &   22.525 &  33.2 \\
090728  & 14:45:45 &  270485147 &  29.644 &   41.632 &  60.6 \\
090813  & 04:10:43 &  271829445 & 225.065 &   88.571 &  34.9 \\
090831C & 21:30:25 &  273447027 & 108.294 &  -25.112 &  41.0 \\
090915  & 15:35:36 &  274721738 & 237.990 &   15.480 &  72.7 \\
091010  & 02:43:09 &  276835391 & 298.669 &  -22.538 &  55.4 \\
091127  & 23:25:45 &  281057147 &  36.581 &  -18.948 &  25.8 \\
091202  & 23:10:12 &  281488214 & 138.831 &   62.544 &  23.5 \\
091221  & 20:52:52 &  283121574 &  55.798 &   23.243 &  54.5 \\
091230  & 06:27:30 &  283847252 & 132.915 &  -53.882 &  31.9 \\
100111A & 04:12:49 &  284875971 & 247.029 &   15.539 &  32.1 \\
100203A & 18:31:07 &  286914669 &  96.225 &    4.793 &  32.3 \\
100206A & 13:30:05 &  287155807 &  47.168 &   13.175 &  44.4 \\
100212A & 14:07:22 &  287676444 & 356.445 &   49.492 &  20.9 \\
100316D & 12:44:50 &  290436292 & 107.599 &  -56.275 &  50.5 \\
100401A & 07:07:31 &  291798453 & 290.813 &   -8.257 &  44.7 \\
100418A & 21:10:08 &  293317810 & 256.358 &   11.457 &  64.5 \\
100427A & 08:31:55 &  294049917 &  89.171 &   -3.461 &  29.7 \\
100514A & 18:53:58 &  295556040 & 328.821 &   29.170 &  60.6 \\
100528A & 01:48:05 &  296704087 & 311.119 &   27.810 &  49.6 \\
100614A & 21:38:26 &  298244308 & 263.534 &   49.232 &  37.8 \\
100704A & 03:35:08 &  299907310 & 133.639 &  -24.202 &  63.1 \\
100719A & 03:30:57 &  301203059 & 112.319 &   -5.857 &  31.3 \\
100725B & 11:24:34 &  301749876 & 290.029 &   76.955 &  24.0 \\
100728A & 02:18:24 &  301976306 &  88.753 &  -15.259 &  59.1 \\
100902A & 19:31:54 &  305148716 &  48.626 &   30.970 &  55.1 \\
101129A & 15:39:31 &  312737973 & 155.921 &  -17.645 &  25.3 \\ 101219B & 16:27:53 &  314468875 &  12.259 &  -34.556 &  52.0 \\
101224A & 05:27:13 &  314861235 & 285.939 &   45.706 &  73.3 \\
110102A & 18:52:25 &  315687147 & 245.877 &    7.617 &  37.7 \\
110119A & 22:20:58 &  317168460 & 348.589 &    5.982 &  71.9 \\
110128A & 01:44:33 &  317871875 & 193.871 &   28.108 &  45.1 \\
110207A & 11:17:20 &  318770242 &  12.540 &  -10.790 &  71.6 \\
110223A & 20:56:59 &  320187421 & 345.386 &   87.586 &  44.0 \\
110328A & 12:57:45 &  323009867 & 251.233 &   57.590 &  46.6 \\
110411A & 19:34:11 &  324243253 & 291.427 &   67.706 &  33.3 \\
110412A & 07:33:21 &  324286403 & 133.491 &   13.488 &  54.3 \\
110414A & 07:42:14 &  324459736 &  97.876 &   24.349 &  18.3 \\
110610A & 15:21:32 &  329412094 & 308.205 &   74.827 &  66.2 \\
110903A & 02:39:55 &  336710397 & 197.061 &   58.985 &  44.4 \\
111029A & 09:44:40 &  341574282 &  44.785 &   57.101 &  15.5 \\
111208A & 08:28:11 &  345025693 & 290.215 &   40.669 &  47.7 \\
120212A & 09:11:22 &  350730684 &  43.086 &  -18.043 &  48.8 \\
120215A & 00:41:15 &  350959277 &  30.057 &    8.790 &  54.0 \\
\enddata
\end{deluxetable}

\subsection{Matched filter weight computation and significance}
The heart of our signal detection technique is the use of a matched filter to optimize the probability for correctly identifying a GRB in the presence of random backgrounds of gamma-rays. The matched filter technique is widely employed for time domain detection of radar and sonar signals but can be easily extended to more complex signal structures. In the simplest case, assume a uniform background noise, $n(t)$, and a signal waveform, $s(t)$. The detection significance for the signal is maximized by integrating a matched filter, $f(t)$, over the effective duration of the signal. Variational methods show that $f(t) = c \cdot s(t)$ where $c$ is a constant. For situations with non-constant backgrounds, the appropriate choice for $f(t)$ is given by the ratio, $c \cdot s(t)/n(t)$. As described in Akerlof et al. 2011, the total weight for each photon was computed by the product of four quantities representing the estimated signal to background ratio for energy, position on the sky, time of arrival and LAT photon detection class. Finally, a somewhat ad hoc weight sharing factor, $\zeta$, was applied to kill events in which the apparent total event weight was heavily dominated by only one or two photons, thus favoring the extreme tails of the background distribution.

The photon data selection criteria for all events is the same as described in our previous work (Akerlof et al. 2010, 2011 and Zheng et al. 2012b, 2012c). Briefly stated, the photon energy is restricted to the range from 100 MeV to 300 GeV. A zenith angle cut of less than 105$^\circ$ is applied as recommended by the LAT team. A detection time window is defined by the interval from T0 to T0+47.5s where T0 is the nominal trigger time. The photons passing these three criteria are then used for further analysis. The matched weight for each photon was calculated by the method described in Akerlof et al. 2011, equations 1 - 5. Note that all classes of photons are included together in computing the matched weight score using our previously described methodology. A weight-sharing factor, ${\zeta}$, was computed via equation 8 and combined to yield the quantity, $\zeta \sum w_i$. The final result is the sum of the matched weights for all single photons multiplied by the weight-sharing factor, ${\zeta}$. It is this value, ${\zeta}$, that is propagated to all further analysis. Note that for each of the four co-add sets described above, the photons for every matching LAT data set were rotated to a common axis and the co-added ensemble was handled as if it were a single GRB event (see Figure 1 for a sky map of the "$Swift\_$GBM" co-add composite field).

These various considerations determined how we treated two GRBs with previously reported LAT detections, GRB100728A and GRB110625A. GRB 100728A (Abdo et al., 2011) has been included in our analysis since although LAT emission was found to be associated with a late time X-ray flare, no LAT emission was reported during the GRB prompt phase from 0 to 167 s. In the case of GRB 110625A (Tam et al., 2012), the event was excluded on the basis of the LAT boresight angle of 88$^\circ$ when the GRB occurred.

The first question is whether there is evidence for any significant association of high energy photons with these GRB fields. To test this assumption, 1256 sets of random co-added fields were constructed with LAT data appropriate for each of the four samples with 79, 46, 14 and 32 GRB events respectively. For each of the 1256 random co-added sets, every single constituent field was selected at random time but with center position identical to the corresponding GRB field. These random fields also had boresight and zenith angles similar to the corresponding GRB data. The distribution of these weights is shown in Figure 2 for the "$Swift\_$GBM" sample of 46. In this case, 7 random fields out of the 1256 yielded a weight greater than for the actual set associated with GRBs. Similar results were obtained for the other samples as well, setting an overall confidence level in the neighborhood of 99\% that a modest excess of GRB-related photons has been detected. A synopsis of these estimates is provided in Table 2.

The conclusion that a few GRB photons were associated with Swift-triggered events was probed in two different ways. It was noted that the fields associated with the 79 GRB events were not uniform in intensity as measured by the sum of matched filter weights over the 1256 observations of each field taken randomly in time. By comparison of one half of the random data observations with the other, a substantial pair-wise correlation was observed for each individual GRB direction. The noisiest 8 GRB fields were removed and the previous analysis re-run to see if background astrophysical sources were a major contamination. The significance levels for the GRB signals compared to the random co-added sets remained at the $\sim$99\% level for all except the "$Swift\_$dim$\_$GBM" sample. Secondly, it was noted that the matched filter weight method was moderately sensitive to infrequent statistical fluctuations with large weight values. To avoid this possible instability, a ranking procedure was invoked that involved two separate steps. First, the 80th percentile weight was determined for every set of 1256 random fields associated with each GRB event. Next, a score was computed for every random and GRB field based on simply summing the number of two highest photon weights above the 80th percentile cut for each field. For the two co-add samples with the greatest GBM fluences, "$Swift\_$GBM" and "$Swift\_$bright$\_$GBM", there appeared to be a significant correlation with the GRB co-add sets surpassing 93\% and 97\% of the equivalent random groups. In summary, we believe that this analysis has found reasonable evidence for a few GRB photons within a number of events but this number is remarkably small.

\begin{figure}[!]
\centering
   \plotone{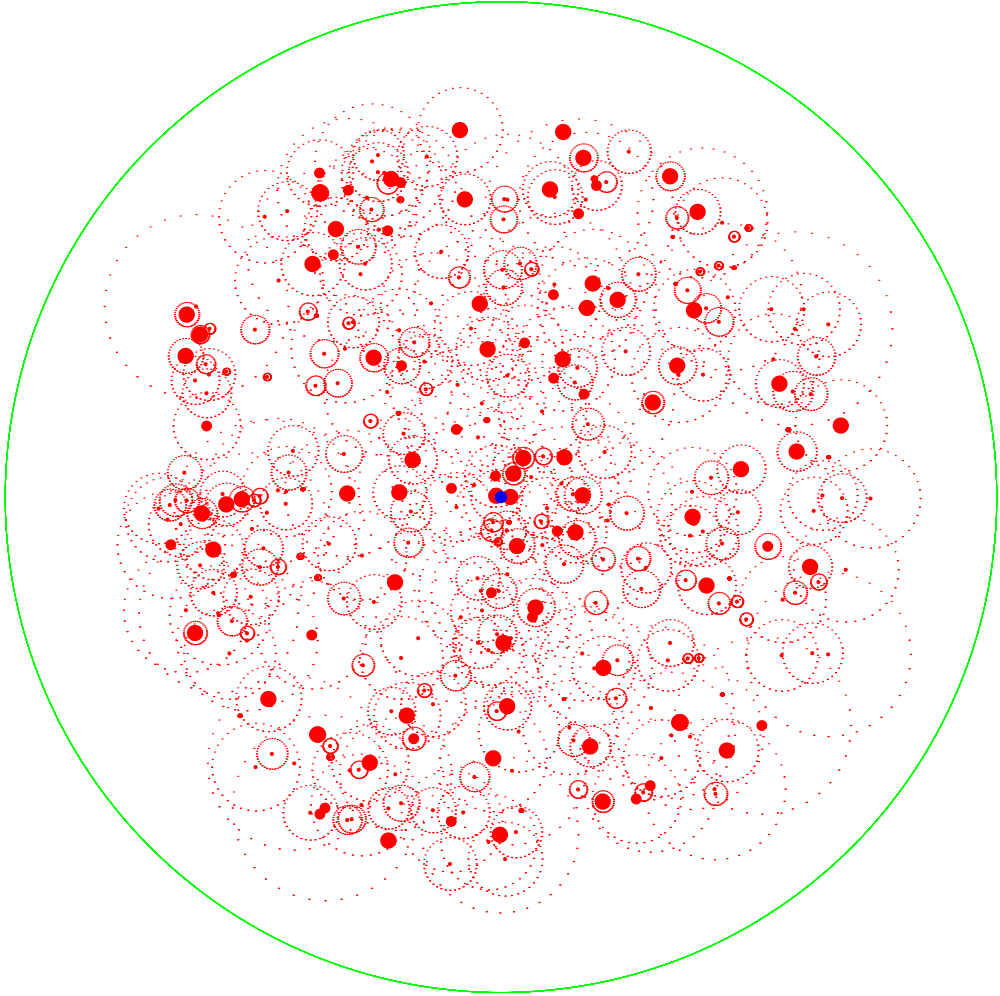}
   \caption{LAT high energy photon sky map for the photons created by co-adding 46 GRB fields triggered by Swift and confirmed by the Fermi GBM. Each field is centered on the GRB direction determined by Swift which is indicated by the blue dot. The relative celestial location of each photon is shown by a red dot whose size indicates the LAT photon event class (class 1 = small dot, class 2 = medium dot, class 3 or 4 = large dot)(Atwood et al. 2009). The photon angular PSFs, estimated from the energy, are indicated by the surrounding dotted circles. The green boundary circle with a radius of 16.0$^\circ$ provides an angular scale. The plot axes are aligned so that North is up and East is to the right. \label{lightcurve}}
\end{figure}

\subsection{Estimating the number of GRB high energy photons}
Since the Swift sample is apparently associated with a non-zero number of GRB photons, the next step was to modify the 1256 random fields previously described by adding a fixed number of GRB photons taken randomly, with replacement, from the set of 851 photons described in Table 2 of Zheng et al. 2012c. Since the only parameter of interest is the GRB photon weight, these values were chosen by randomly selecting values from the appropriate 851-fold array. This procedure was iterated 1000 times for a total of 1256000 co-added fields for each integer number, $n$, of injected GRB photons. The integral distributions of the matched filter weights for the ensemble are shown in Figure 2 for pure random background fields (red line) and 1 to 20 injected GRB photons (blue lines). To find the most probable photon number for each data set, the integral distribution for each curve was estimated at the actual matched filter value. The quoted photon number was interpolated from the two values that bracketed 50\%. A similar procedure was pursued for the $\pm 1$-$\sigma$ ranges by finding the crossing points where the integral distributions took on the values of 0.158655 and 0.841345. The results for each of the four GRB sample sets are listed in Table 2. In order of increasingly more stringent GBM fluence requirements, the average number of photons per unit area is (0.23, 0.31, 0.53) m$^{-2}$, consistent with the high-energy/low-energy correlation shown in Figure 5 of Zheng et al. 2012c.
(The "$Swift\_$dim$\_$GBM" field has a corresponding flux of 0.30 photons m$^{-2}$.) This fluence trend lends additional support to the evidence of a small but finite number of GRB photons in the $Swift$-triggered co-added fields.

\begin{deluxetable}{crrrrrrr}
 \tabcolsep 0.4mm
 \tablewidth{0pt}
 \tablecaption{Summary of Swift-triggered co-added fields}
  \tablehead{\colhead{data set} & \colhead{number of} & \colhead{matched} & \colhead{estimated} &  \colhead{photon \#} & \colhead{effective$^a$} & \colhead{GBM$^b$} \\  \colhead{} & \colhead{co-added} & \colhead{filter} & \colhead{photon} & \colhead{$\pm 1$-$\sigma$} & \colhead{area} & \colhead{fluence} \\  \colhead{} & \colhead{events} & \colhead{weight} & \colhead{number} & \colhead{range} & \colhead{(m$^2$)} & \colhead{($\mu$erg cm$^{-2}$)}} \startdata
All$\_Swift$           & 79 & 587.4 & 8.52 & 4.54 - 13.54   & 36.23 &\\
$Swift\_$GBM           & 46 & 370.0 & 6.15 & 3.15 - 10.15   & 19.89 &  500.7 \\    
$Swift\_$bright$\_$GBM & 14 & 111.8 & 3.21 & 1.54 - ~5.60   & 5.96  &  396.7 \\
$Swift\_$dim$\_$GBM    & 32 & 228.9 & 4.33 & 2.03 - ~7.51   & 13.93 &  104.0 \\
\enddata
\tablenotetext{a}{Sum of the individual effective areas of all GRBs in the sample. The areas are determined from the LAT boresight angles at the burst trigger time.}
\tablenotetext{b}{Sum of the GBM fluences of all GRBs in the sample. These numbers are obtained from from the Fermi GBM Burst Catalog, FERMIGBRST (http://heasarc.gsfc.nasa.gov/W3Browse/all/fermigbrst.html), and the GCN catalog (searchable via GRBlog at http://grblog.org/grblog.php).}
\end{deluxetable}

\begin{figure}[!hbp]
\centering
   \plotone{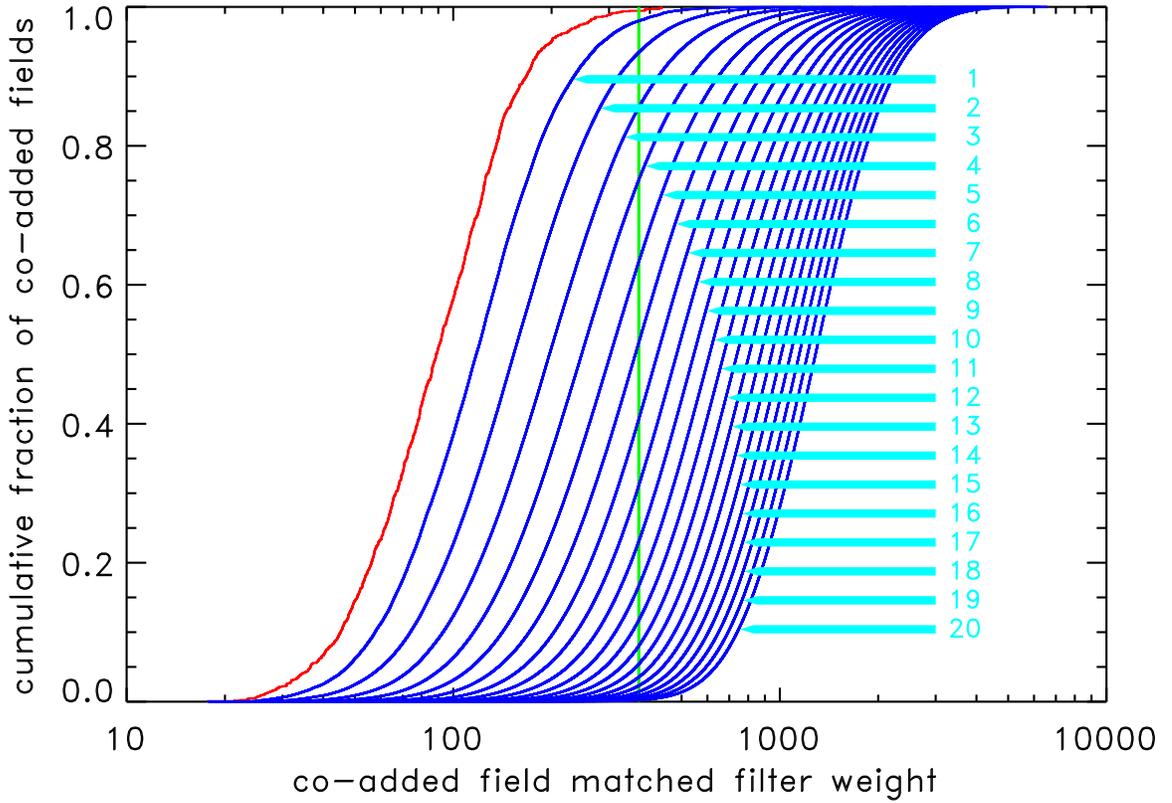}
   \caption{Cumulative distributions of LAT random fields with 0 to 20 injected GRB photons as a function of matched filter weight. The red line shows the distribution of matched weights for 1256 fields obtained by co-adding 46 LAT images but no additional simulated photons.  The blue lines show similar distributions for 1256000 simulated fields with 1 to 20 photons added to each. The vertical green line indicates the value for the "Swift$\_$GBM" co-add field. \label{skymap}}
\end{figure}

\section{Implications for the number distribution of high energy GRB photons}
In a recently completed paper, we have found that the distribution of GRBs with photons $>100$ MeV can be characterized by a distribution function that is proportional to the fixed power of the number of photons per unit area (Zheng et al. 2012c). This result is shown in Figure 3 which depicts the distribution of actual GRB events, the parent distribution and the distribution modified by the effective area of the LAT detector and analysis event discrimination efficiencies. As noted, the inferred number of possible GRB photons in the Swift co-added fields is relatively small. Thus for comparison with observation, we can try making the assumption that the GRB number distribution can be extrapolated below the detection threshold that originally defined the fit. We find that if the event distribution function observed at higher LAT fluxes can be extrapolated to lower rates, the inclusion of a fraction of all GRBs is sufficient to accommodate the numbers inferred from our analysis. Pursuing such questions will eventually offer a better insight as to whether high energy photon emission is a generic feature of all bursts or only a special subset.

\subsection{Computing total photon number from a power law distribution}
We need to find an expression for the total number of photons that can be expected from a set of bursts that are distributed in photon number according to a power law. Assume the number of GRB events emitting $n$ photons per unit area is given by 
\begin{equation}    \label{eq:PL}
\frac{dN}{dn} = c\cdot n^p
\end{equation}

The exponent, $p$, is approximately -1.8, leading to the conclusion that this mathematical form is only valid above some minimum threshold value for $n$ which will be designated as $n_{min}$. Normalizing the integral of $dN/dn$ to unity defines the maximum range over which the power law model can operate:
\begin{equation}    \label{eq:PL}
n_{min} = (- \frac{c}{p+1})^{\frac{1}{p+1}} \end{equation}

For the problem at hand, we need to compute the total number of photons, $\nu_{Swift}$, that can be expected in the co-added field corresponding to all $Swift$ triggers within the co-added set. Each GRB event will be characterized by a specific effective area determined by the boresight angle of the GRB with respect to the LAT z-axis. The probability of contributing $m$ photons to the co-added field from the $i$'th GRB event is determined by the Poisson distribution:
\begin{equation}    \label{eq:PL}
p_i(m) = \int^{\infty}_{n_{min}} {\frac{(n\cdot a_i)^m}{m!} e^{-n\cdot a_i} \frac{dN}{dn} dn} \end{equation}

where $a_i$ is the LAT effective area appropriate for the $i$'th GRB event. The total number of photons is then given by
\begin{equation}    \label{eq:PL}
\nu_{Swift} = \sum_{m=1}^{\infty} m\cdot\eta (m) \sum_{i=1}^{N_{GRB}}
p_i(m)
\end{equation} where $\eta (m)$ is the probability that an event with $m$ photons would escape normal detection criteria. In practice, this limits the summation over  to 8 photons or less. $\eta (m)$ was computed by first estimating the probability that an event with $m$ photons would have been previously detected by the matched filter weight method described in our earlier papers. This corresponds to a matched filter weight sum of approximately 15.0. An essentially pure sample of high energy GRB photons was obtained from 4 bright GRBs, 080916C, 090510, 090902B and 090926A with 125, 176, 164 and 177 photons respectively. For each of these burst events, a Monte Carlo program repeatedly computed the matched filter weight sum for $m$ randomly selected photons to obtain an average probability of exceeding the threshold of 15.0. The independent results for the photon samples from the four bright GRBs were substantially similar and the average for each value of $m$ estimates $\eta '(m)$, the probability of detection by previous searches. $\eta (m)$ is computed immediately as $1-\eta '(m)$.

Since $dN/dn$ is simply a power of $n$, the evaluation of $\nu_{Swift}$ devolves immediately to a summation of terms, each of which can be represented by an incomplete gamma function defined by
\begin{equation}    \label{eq:PL}
\Gamma (z,x) = \int_x^{\infty} t^{z-1} e^{-t} dt \end{equation}

Such functions can be found on familiar mathematical packages such as IDL and Mathematica. Taken together, equations 4 \& 5 link assumptions about the GRB event photon number distribution to the actual number of photons that should be observed for a specified ensemble of exposures. In this way, the results given in Table 2 will lead to limits for these otherwise inaccessible regions of the event photon number distribution.

\begin{figure}[!hbp]
\centering
   \plotone{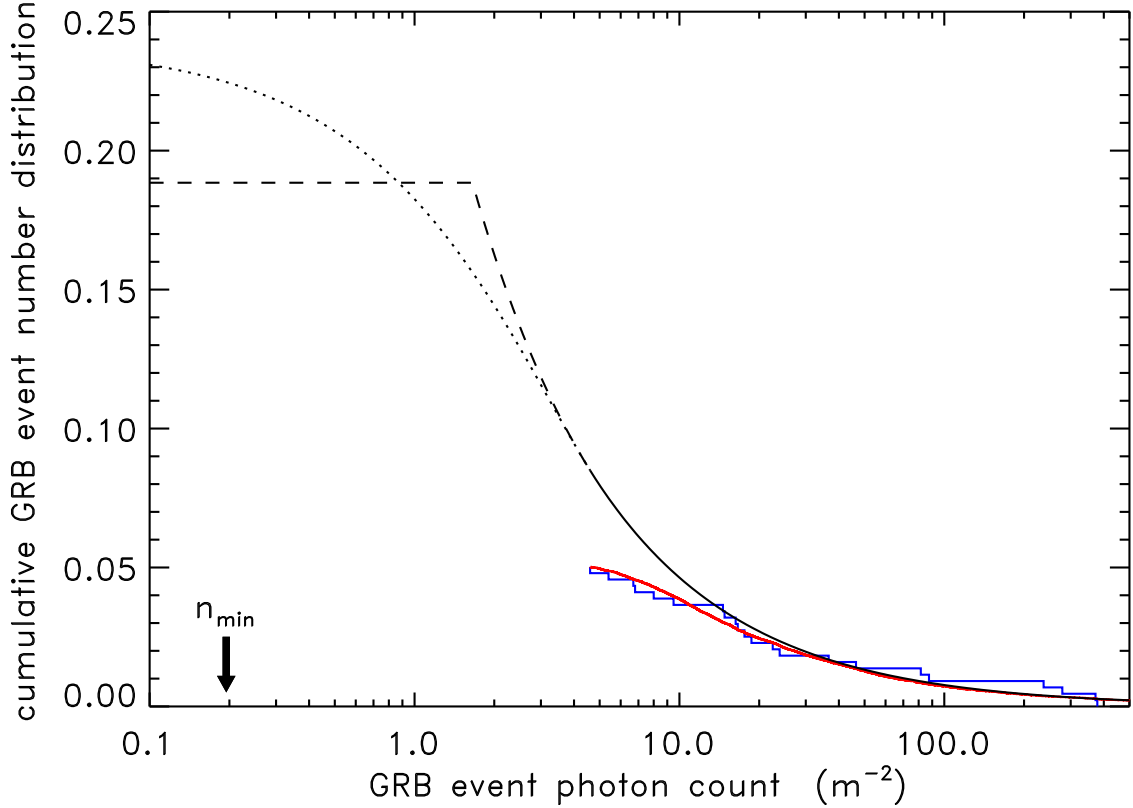}
   \caption{Complement of the cumulative GRB event number distribution, $N(n)$, as a function of the LAT high energy photon flux, $n$. The blue histogram shows the cumulative distribution for the LAT-detected events listed in Table 2 of Zheng et al. 2012c. The red curve is the fit to a power-law corrected for detection efficiencies. The black lines show the underlying power-law distribution in the region of the fit (solid line), extrapolated to satisfy the number of observed sub-threshold quanta (dashed line) and a similarly constrained power-law segment with a slightly greater number of events with lower fluxes (dotted line). The arrow marked $n_{min}$ indicates where an extrapolation of the power law distribution to lower values of the photon flux, $n$, would reach unity, ie. include all GRBs. \label{skymap}}
\end{figure}

\subsection{Reconciling the Swift-triggered photons and the extrapolated number distribution}
As noted previously, the lower limit of the range for a power-law distribution with negative exponent is bounded by the finite number of events it describes. That condition is satisfied by setting the lower bound, $n_{min}$, by equation (2). Thus, the obvious first question is how many photons would be expected in the "Swift$\_$GBM" co-add field if all 46 events obeyed this distribution over the range from $n_{min}$ to $\infty$. With $n_{min}$ defined by equation 2, this can be obtained by application of the mathematical relations of equations 4 \& 5. For the value of $p$ which best fits the observed event distribution shown in Figure 3, the presumption that all GRBs contribute is clearly violated at the 97\% confidence level as shown in Figure 2 by overpredicting the number of photons, 14, instead of the observed value of 6 given in Table 2.

The obvious next step is to inquire where the lower bound should be set to match the observed photon number. To keep the notation unambiguous, this bound will be designated $n_{thresh}$ and $n_{thresh} > n_{min}$. For the most probable values for $p$ and $\nu_{Swift}$, the value for $n_{thresh}$ is 1.7 photons/m$^2$ corresponding to 19\% of the full GRB sample ($n_{min} = 0.2$ photons/m$^2$). Varying $\nu_{Swift}$ over the $\pm$1-$\sigma$ range changes the percentage of the contributing GRBs from 11\% to 40\%. Keeping $\nu_{Swift}$ at its central value but varying $p$ over its $\pm$1-$\sigma$ range changes these percentages by a factor of roughly two. However if $p$ and $\nu_{Swift}$ are simultaneously set to their 1-$\sigma $   upper limits of -1.5 and 10, the fraction of GRBs rises to 100\%. Admittedly, this extrapolation from present observations relies on extrapolation of the photon number distribution to low photon counts. Thus, the analysis should be viewed with circumspection but is intended to illustrate the twin constraints of an apparent power-law distribution and the limited total number of detected photons associated with a considerable number of bursts.

The abrupt flux cutoff event distribution model described above is an unnatural extreme that favors a few events with fluxes near the current lower detection threshold. An alternative is to model the differential distribution in the region from zero to the detection threshold as a sum of two terms chosen to match the fitted distribution at threshold. With the constraint implied by the total of number of events inferred from the co-added analysis, the cumulative distribution plateaus at 24\% of all GRBs (see dotted curve in Figure 3). The differential distribution for this segment is of the form, $a+bn^q$, where $a$ and $b$ are constants, $n$ is the GRB event flux and $q$ is a fractional exponent chosen to match the co-added field photon count. This tends to favor a larger number of events at lower fluxes. Thus, despite wide differences in model assumptions, the GRB event population seems to divide between the $\sim$25\% that emit high energy radiation and the 75\% that don't. We find it exceedingly strange that nature has found a way to limit the number of events with high energy flux intensities in the 1 to 4 photons/m$^2$ range relative to what occurs at higher fluxes. Addressing this question might lead to a better understanding of the GRB high energy emission process. Finally, the stringent limits for total photon numbers in the various co-added trigger fields strongly suggest that there are very few modest intensity bursts that have gone undetected because of inadequate signal processing techniques. 

\section{Summary}
In this paper, we have found a small but statistically significant number of GRB-associated photons for co-added sets of events triggered by the Swift instrument. Although this number has large statistical errors, it points to a conclusion that high energy photon emission is relatively infrequent and can not be hidden in the somewhat fainter burst events. The necessary initial conditions that give rise to this phenomenon are not at all understood. With a larger event sample, it may be possible to better classify the population of high energy emitting GRBs and test models that explain their behavior. If nothing else, this analysis again demonstrates the importance of GRB detection with good angular resolution.

\acknowledgments
We thank Timothy McKay for constructive suggestions for this manuscript. This research is supported by the NASA grant NNX08AV63G and the NSF grant PHY-0801007.

\end{document}